\documentclass[12pt]{article}      

\usepackage{caption}
\usepackage{fancyhdr}
\usepackage{longtable}
\usepackage{tabu}
\usepackage[section]{placeins}   %
\usepackage[yyyymmdd,hhmmss]{datetime}
\usepackage[pagebackref]{hyperref}  
\hypersetup{ colorlinks,
    citecolor=blue,
    filecolor=blue,
    linkcolor=blue,
    urlcolor=blue
}
\renewcommand*{\backref}[1]{}

\newcommand{\captionl}[1]{\captionsetup{width=.9\textwidth}{\caption[dummy]{{\sf #1}}}}

\usepackage{xcolor}

\definecolor{gray}{rgb}{0.6,0.6,0.6}
\definecolor{red}{rgb}{0.85,0,0}
\definecolor{green}{rgb}{0,0.85,0}
\definecolor{blue}{rgb}{0,0,0.85}
\definecolor{beige}{rgb}{0.92,0.87,0.78}
\usepackage[all]{hypcap}    

\usepackage{overcite}       
\usepackage{graphicx}
\graphicspath{ {./Figures/} }

\makeatletter \renewcommand\@biblabel[1]{$^{#1}$} \makeatother
 \newlength{\bibhang}
\newcommand{\cen}[1]{\begin{center} #1 \end{center}}
\newcommand{\eqn}[1]{\begin{equation} #1 \end{equation} }

\newcommand{\ie}{{\it i.e.}, }
\newcommand{\eg}{{\it e.g.}, }

\newcommand{\io}{{${}^{125}$I }}
\newcommand{\pa}{{${}^{103}$Pd }}
\newcommand{\ioc}{{${}^{125}$I}}
\newcommand{\pac}{{${}^{103}$Pd}}

\newcommand{\cs}{{${}^{131}$Cs }}
\newcommand{\cscc}{{${}^{131}$Cs}}

\newcommand{\pd}{\pa}

\newcommand{\eb}{{\tt egs\_brachy }}
\newcommand{\ebc}{{\tt egs\_brachy}}
\newcommand{\g}{{\tt g }}

\newcommand{\BD}{{\tt BrachyDose }}
\newcommand{\BDc}{{\tt BrachyDose}}

\begin{document}

\cen{{\Large \bfseries Update  of the CLRP TG-43 parameter database for low-energy
brachytherapy sources}} 
\vspace{5mm}

\pagestyle{empty}
\pagenumbering{roman}
\cen{Habib Safigholi, Marc~J.~P.~Chamberland$^{a)}$, Randle~E.~P.~Taylor$^{b)}$,
    Christian~H.~Allen, Martin~P.~Martinov, D.~W.~O.~Rogers,
    and Rowan~M.~Thomson \\
Carleton Laboratory for Radiotherapy Physics (CLRP), Department of Physics,\\ 
Carleton University, Ottawa, Ontario, K1S 5B6, Canada\\\vspace{2mm}
\noindent \begin{small}$^{a)}$ Present address, Medical Physics, The University of
Vermont Medical Center, Burlington, Vermont, 05401, USA\\
$^{b)}$ Present address, Multi Leaf Consulting, Port Elgin, ON, Canada, N0H
2C3 \end{small}

}
\begin{abstract}
\noindent{\bf Purpose:} To update the 
Carleton Laboratory for Radiotherapy Physics (CLRP) TG-43 dosimetry database for
low-energy ($\leq50$~keV) photon-emitting low-dose rate (LDR) brachytherapy sources utilizing
the open-source EGSnrc application \eb rather than the \BD application used
previously for 27 LDR sources in the 2008 CLRP version (CLRPv1). 
CLRPv2 covers 40 sources (\pac,  \ioc, and \cscc). A comprehensive set of
TG-43 parameters is calculated, including dose-rate constants, radial dose
functions with functional fitting parameters, 1D and 2D anisotropy
functions, along-away dose-rate tables, Primary-Scatter separation dose
tables (for some sources), and mean photon energies at the surface of
the sources. 
The database
also documents the source models which will become part of
the \eb distribution.\\
{\bf Acquisition and Validation Methods:} Datasets are calculated after a
systematic recoding of the source geometries using the
egs++ geometry package and its \eb extensions.
Air-kerma strength per history is calculated for
models of NIST's Wide-Angle Free-Air chamber (WAFAC) and for a point detector
located at 10~cm on the source's transverse axis. Full scatter  water phantoms
with varying voxel resolutions in cylindrical coordinates are used for dose
calculations. New statistical uncertainties of source volume corrections
for phantom voxels which overlap with brachytherapy sources are implemented
in \ebc, and all CLRPv2 data include these uncertainties. 
For validation, data are compared to CLRPv1 and other data in the literature. 
\\ 
{\bf Data Format and Access:}
Data are available 
at \url{https://physics.carleton.ca/clrp/egs_brachy/seed_database_v2}. As
well as being presented graphically in comparisons to previous calculations, data
are available in Excel (.xlsx) spreadsheets for each
source. \\
{\bf Potential Applications:} The database 
has applications in 
research, dosimetry, and brachytherapy planning. This comprehensive update
provides the medical physics community with more accurate TG-43 dose
evaluation parameters, as well as fully-benchmarked and described
source models which are distributed with \ebc. 
\mbox{~}\vspace{-11mm}\\
\end{abstract}
\vspace{2mm}\noindent {\bf Key words:} Low-energy brachytherapy, CLRP, TG-43, database, 
EGSnrc, Monte Carlo \ebc
 \clearpage

\vfill
\newpage

\setcounter{page}{1}

\setlength{\baselineskip}{0.7cm}

\pagestyle{fancy}
\pagenumbering{arabic}
\section{Introduction}\label{intro}

The TG-43 protocol\cite{Na95} presented a worldwide standard dosimetry
formalism for brachytherapy sources and has led to an estimated 5\%
uncertainty on brachytherapy dosimetry\cite{tg138}.  Presently,
brachytherapy dosimetry and planning is widely  based on the single source
consensus data published in reports:
TG-43 (1995)~\cite{Na95}, the 2004 update (TG-43U1)~\cite{Ri04,Ri04a},
the supplements in 2007 (TG-43U1S1)~\cite{Ri07a} and 2017
(TG-43U1S2)~\cite{Ri17,Ri18}, and the AAPM/ESTRO TG-152 report for
High-Energy Brachytherapy source Dosimetry (HEBD)~\cite{Pe12}. Monte Carlo
(MC) calculations are the predominant method of determining consensus
values for the radial dose function, $g(r)$, and the 2D anisotropy
function, $F(r,\theta)$, for both low- and high-energy sources. Consensus
dose-rate constants (DRCs or $\Lambda$) for low-energy sources are
currently taken as
averages of the mean of MC calculations and the mean of measurements
since historically there was a systematic difference between the two (see
ref~\cite{RR14} for a summary). In contrast, consensus DRCs
for high-energy sources are based solely on MC calculations\cite{Pe12}.

In 2008, version 1 of the \textbf{C}arleton \textbf{L}aboratory for
\textbf{R}adiotherapy \textbf{P}hysics TG-43 database  (CLRPv1) was
published.~\cite{TR08b,TR08c,TG43WEB}. It was compiled using the EGSnrc MC
application \BD and contains DRCs, $g(r)$, and $F(r,\theta)$~values for 42
low- and high-energy brachytherapy sources.  These data were extensively
used in the TG-43U1S2 report~\cite{Ri17,Ri18}. Since the 2008 CLRPv1
version of the TG-43 database, several low- and high-energy photon-emitting
brachytherapy sources have become available and are on the AAPM/IROC
Houston Registry of Brachytherapy Sources
(\url{http://rpc.mdanderson.org/rpc/}). Also, various minor errors in
modelling the source geometries and in MC simulations reported in CLRPv1
have been uncovered. Furthermore, a fast, versatile, and open source
EGSnrc application, called \ebc, was developed within the CLRP~\cite{Ch16}
and recently released as open-source software.~\cite{Th18}

The goal of the present work is to perform a comprehensive update of the
CLRPv1 TG-43 database parameters for low-energy, low-dose rate (LDR)
brachytherapy sources with a single consistent approach using \ebc. This
``CLRPv2'' update includes an independent recalculation of the datasets
with more precision and is based on the egs$++$ package (Kawrakow et al.,
2018\cite{Ka18}) with some extensions created for the \eb
application\cite{Ch16}.  The CLRPv2 dataset includes a total of 40
low-energy sources for radionuclides \pac, \ioc, and \cscc; higher energy
brachytherapy sources will be added in future work.   Some of these sources
are no longer in production; however, they are included to provide
dosimetry parameters for retrospective analyses and research. The
brachytherapy source geometry models described and benchmarked by the
comparisons in the CLRPv2 database will be released for use with the open
source \eb application.

\section{Acquisition and Validation Methods}\label{methods}

\subsection{Computational tools and Monte Carlo simulations}
\label{methods:mc} 
All calculations are performed with the EGSnrc application \ebc\cite{Ch16}
(GitHub commit hash 4f3ecac version with additions regarding the
statistical uncertainties in the volume corrections described below).
This is free and open-source software for doing rapid MC
brachytherapy dose calculations available via
\url{https://physics.carleton.ca/clrp/egs_brachy/}.  Electron transport,
although available in \ebc, is not modelled in the current calculations.
The photon cutoff energy is set to 1~keV, except for air-kerma strength
calculations for which
the photon cutoff energy is 5~keV to eliminate the characteristic x rays
from Ti as required by the definition\cite{Ri04}.  Rayleigh scattering,
bound Compton scattering, photoelectric absorption, and fluorescent
emission of characteristic x rays are all simulated. Photon cross sections
are taken from the XCOM database~\cite{BH87} and mass energy absorption
coefficients are calculated with the application \g  as distributed with
the EGSnrc package before 2017. Recent improvements\cite{RT19} in the
EGSnrc code and the application \g mean that these mass energy absorption
coefficient values would change by up to a maximum of 0.2\% using the newer
releases of EGSnrc.  There is ambiguity in whether renormalized or
unrenormalized Scofield photoelectric cross sections are in better
agreement with experimental data \cite{ICRU90}; the current work employs
the unrenormalized cross sections (consistent with EGSnrc default
\cite{Ka18}).  Initial photon energies and probabilities are sampled from
the NNDC~\cite{NNDC} spectra for $^{103}$Pd, and $^{131}$Cs sources. For
\ioc, the photon spectrum from the NCRP Report 58~\cite{NCRP58} is used
since Rodriguez and Rogers~\cite{RR13} demonstrated that using the NCRP
spectrum leads to better agreement with measured spectra for many seeds and
it is consistent with the spectrum recommended by the BIPM for
use by primary standards labs~\cite{BIPM-5-6}.

Dose calculations are done in a full scattering cylindrical water phantom
(30~cm long, 15~cm radius). In addition, for three representative sources,
a larger water phantom is modelled (40~cm long, 20~cm radius).  For
efficiency purposes, collision kerma and hence absorbed dose is scored in
concentric cylindrical shells, as sources are cylindrically symmetric and
there is charged particle equilibrium for low-energy brachytherapy sources.
As discussed previously\cite{Ta06b}, to increase accuracy and reduce bin
size artifacts, the radial ($r$) and length ($z$) resolutions of the
cylindrical shells are 0.1~mm for $r$~$\leq$~1~cm, 0.5~mm for
1~cm~$<$~$r$~$\leq$~5~cm, 1~mm for 5~cm~$<$~$r$~$\leq$~10~cm and 2~mm for
$r>10$~cm where $r$ is the radial distance from the cylinder's axis (and
source is aligned along the $z$-axis).  All calculations employ \ebc's
tracklength scoring of collision kerma.

Air-kerma strength per history,
$S_K^{hist}$, is scored {\em in
vacuo}, using a scoring voxel located on the transverse axis 10~cm away
from the source and of size 0.1$\times$0.1$\times$0.05~cm$^3$ when
approximating a point detector, or 2.66$\times$2.66$\times$0.05~cm$^3$ when
representing the solid angle subtended by the primary collimator of the
NIST WAFAC~\cite{Se03} detector. The two voxel sizes are referred to as
`Point' and `WAFAC' detectors, respectively, on the CLRP TG-43 website.  In
both cases, corrections are made to determine the air-kerma on the axis at
the front of the detector using the same $k_{r^2}$ correction factors as
used for CLRPv1\cite{Ta06b}. The factors used are 1.0050 and 1.0168 for the
point and WAFAC detectors respectively. Strictly speaking, the formula for these
correction factors has been shown to be wrong in general\cite{Ro19}.
However, for the geometries of the detectors used here, they have been
shown to be highly accurate\cite{Ro19}.

For most LDR sources, the point and WAFAC estimates of the air-kerma
strength are the same. However for some low-energy sources in which the
radioactivity is deposited on high-Z radio-opaque surfaces with sharp
corners, the on-axis point estimate is lower because the high-Z materials
attenuate photons from the end surfaces aimed on the axis\cite{Wi00}. This
leads to a well-known dependence on the scoring voxel size when determining
the air-kerma strength\cite{Ta06b} and hence the dose-rate constant.  The
geometry of the detector's sensitive region for air-kerma calculations is
filled with a very low-density dry air, as recommend by TG-43.

Enough histories $(\sim{}5\times10^{10})$ are simulated to ensure that
$g(r)$, and $F(r,\theta)$ results have a k=1 statistical uncertainty $\leq
0.05\%$ at $ r\leq 5~$cm, and DRC results have statistical uncertainties
$\leq$0.2\% for point detectors or $\leq$0.02\% for WAFAC detectors.

\subsubsection{Dealing with long sources}
\label{long_sourcs}
The $k_{r^2}$ corrections required to convert the average kerma in the
WAFAC detector's volume or front surface to the kerma on the axis as required
for determining the air-kerma strength are usually derived analytically assuming
that the source is an on-axis isotropically radiating point source.
However, these assumptions break down for the long CivaTech sources.  
Paxton {\em et al}\cite{Pa08b} showed that the NIST WAFAC cannot, in
principle, directly measure the air-kerma strength for long sources because
some photons would create charge outside the charge-collecting region. This
issue does not affect the University of Wisconsin's (UW's)
variable-aperture free-air chamber (VAFAC)\cite{Pa08b}.  However, that
still leaves the issue of determining the $k_{r^2}$ corrections for
non-point sources. We have done that by calculating the average air kerma
per history in a thin (0.001 cm thick)  circular detector (radius 1.333 cm)
at 10 cm from the axis for a series of point sources off-axis and comparing
that average to the air kerma per history in a small on-axis voxel
($0.1\times 0.1\times0.001$~cm$^3$). The detector subtended the same solid
angle as the NIST WAFAC or UW VAFAC detectors at 30 cm.  The values of
$k_{r^2}$ for line sources is then determined by integrating the
corrections as a function of the point source's distance off-axis.  The
values varied from 1.0094 for a point source to 1.0084 for a 5~cm line
source.  These values agree within 0.05\% with those calculated by Paxton
{\em et al}\cite{Pa08b} of 1.0089 to 1.0087 for the same range of source
lengths.  The point source on axis compares well with the analytic value
for detectors of the same area of 1.0094~(square detector) or 1.0090
(circular detector)\cite{Ro19}. On the assumption that there should be no
difference in the air-kerma strength based on the WAFAC geometry or the
point-source geometry, the ratio of the point to WAFAC geometry raw
calculations for the CivaTech sources can also be used to estimate
$k_{r^2}$ factors; on average, they agree within $(0.02\pm0.02)$\% with the
values calculated based on integration of the point-source calculations.

\subsection{Source Volume Correction  }
\label{methods:volcorr}

\begin{figure}[ht]
\begin{center}
  \includegraphics[scale=0.4]{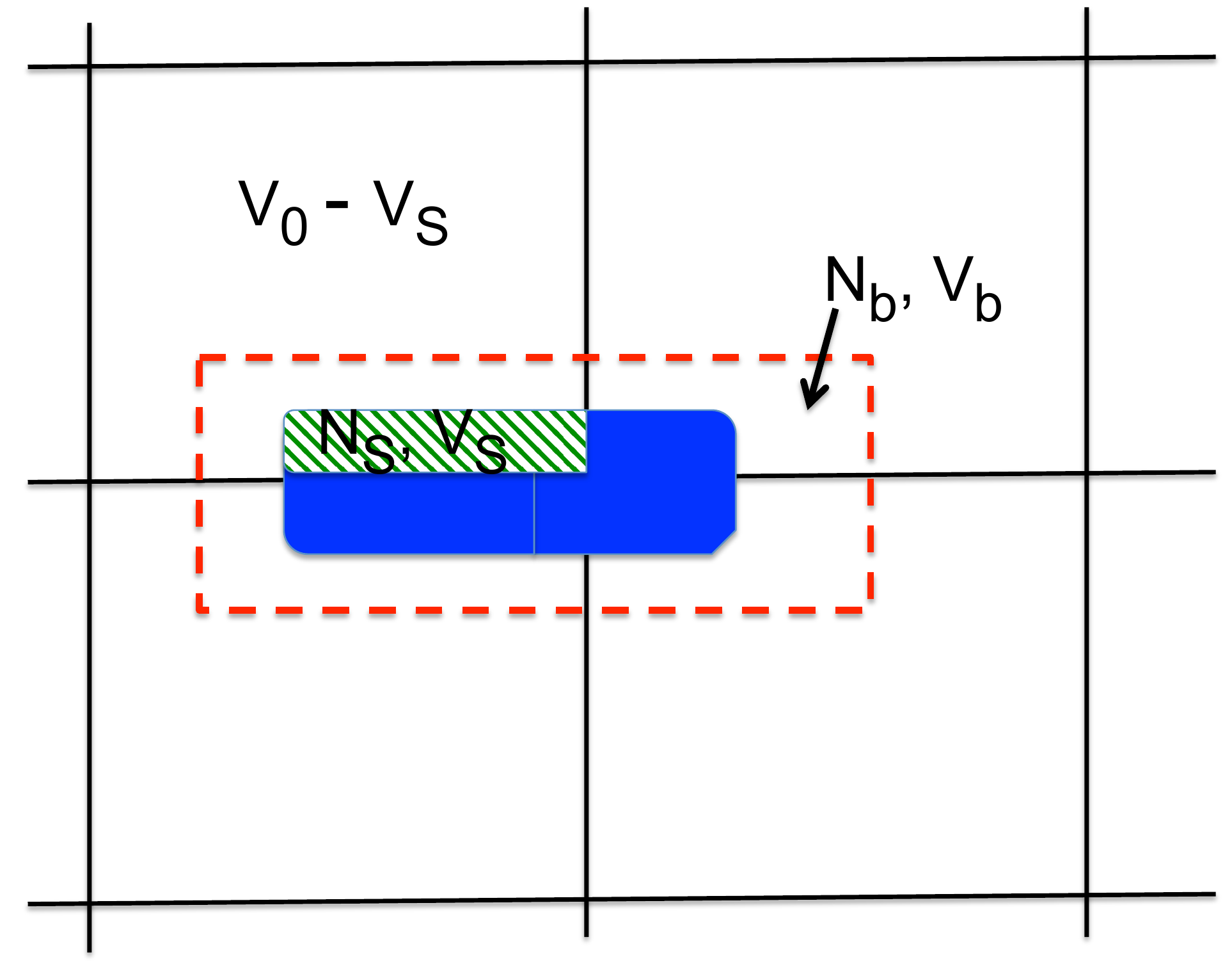}  
\captionl{A schematic diagram of a source volume correction calculation in
\ebc. Volume $V_s$ (hatched region) is the part of the source volume that is
unknown and is needed for volume correction to the voxel with uncorrected
volume $V_0$. The source
boundary shape (red dashes) has volume $V_b$ around the source with $N_b$ random
points generated within the region. 
The number of random points which fall in
both the source and voxel of interest, \ie $V_s$, is $N_s$.
\label{fig_vol_corr} }
\end{center}
\end{figure}

It was observed that dose values in
voxels overlapping a source 
were affected by the sometimes large statistical variations in the
source volume corrections which correct the voxel
volume to account for this overlap. Thus, the statistical uncertainty on
the source volume
correction needs to be quantified and minimized for those
phantom voxels which contain part of the source. 
The source volume correction calculations in
\eb are performed with a MC calculation in one of two source boundary
shapes (cylinder or box) around a brachytherapy source. Random points
within the boundary shape are generated with a specified
random point density (cm$^{-3}$). The number of points which fall
inside both the source and the voxel of interest are tallied 
and also the total number within the boundary shape is known. The
ratio of these two counts, \ie the fraction of the volume of the bounding shape
occupied by the source within the voxel of interest, 
is used to determine the source volume correction. The
density of random points is specified in the input to \eb and a
range from 10$^{3}$ to 10$^{12}$~cm$^{-3}$ can be relevant.
Figure~\ref{fig_vol_corr}  shows a typical source which overlaps 4 phantom
voxels and is surrounded by a boundary shape (dashed line).

For each voxel overlapping the source, \eb calculates the volume $V_s$ by
tracking $N_s$, the number of random points generated in $ V_b$  that fall
in the source region and within the voxel under consideration. Since the
total boundary volume $V_b$ and the total number of points generated by \eb
within the source boundary volume, $N_b$ are known, then the source volume,
$V_s$, is calculated as:
  \eqn{V_s =V_b \cdot \frac{N_s}{N_b} = \frac{N_s}{PD},  \label{eq_Vs}}
where $N_b = PD\cdot V_b$ with $PD$ (cm$^{-3}$) being the density of random points 
used. The corrected volume ($V_{\rm corr}$) is then calculated as:
 \eqn{V_{\rm corr} =V_0 - V_s =V_0-V_b \cdot \frac{N_s}{N_b}. \label{eq_Vcorr}}
As the nature of MC target finding is binary (\ie a point generated in
$V_b$ is either in $V_s$ or not), it can be represented by a discrete
Bernoulli distribution~\cite{Fe67}.  Therefore, if the mean fraction is
$N_s$/$N_b$, the standard deviation of the mean would be
   \eqn{s_{\frac{N_s}{N_b}} = 
    \frac{\sqrt{\frac{N_s}{N_b}(1-\frac{N_s}{N_b})}}{\sqrt{N_b}} = 
    \sqrt{\frac{N_s(N_b-N_s)}{N_b^3}}.
    \label{SD-eq_Vcorr}
    }

Using the error propagation law on the relationship between $V_{corr}$ and
$N_s$/$N_b$ in equation~(\ref{eq_Vcorr}), the final fractional uncertainty
on $V_{corr}$ can be written as:
  \eqn{\left|\frac{s_{V_{corr}}}{V_{corr}}\right|=
  \frac{V_b}{V_0-V_s}\sqrt{\frac{N_s(N_b-N_s)}{N_b^3}}= 
  \frac{1}{(PD\ V_0-N_s)}\sqrt{\frac{N_s(N_b-N_s)}{N_b}}.
  \label{eq_uncertVcorr}
  }
or writing in terms of the volumes involved by using $N_s=V_sPD$ and $N_b
= V_bPD$:
  \eqn{\left|\frac{s_{V_{corr}}}{V_{corr}}\right|=
   \frac{1}{PD(V_0-V_s)}\sqrt{\frac{PD\ V_s(PD\ V_b-PD\ V_s)}{PD\ V_b}}
   =\frac{1}{\sqrt{PD}} \sqrt{\frac{\ V_s(\ V_b - \ V_s)}{V_b(V_0-V_s)^2}} =
\frac{k}{\sqrt{PD}},   \label{eq_vol_only}
  }
where k is a constant depending only on $V_0, V_b$ and $V_s$.

If a source volume correction is needed, its statistical uncertainty can
play an important role and must be included in the statistical uncertainty
on the voxel's dose. Therefore, source volume correction uncertainties have
been implemented in \eb for voxels containing all or part of a source. The
random point density,$PD$, is a user input.

\begin{figure}[ht] 
\begin{center}
\includegraphics[scale=0.54]{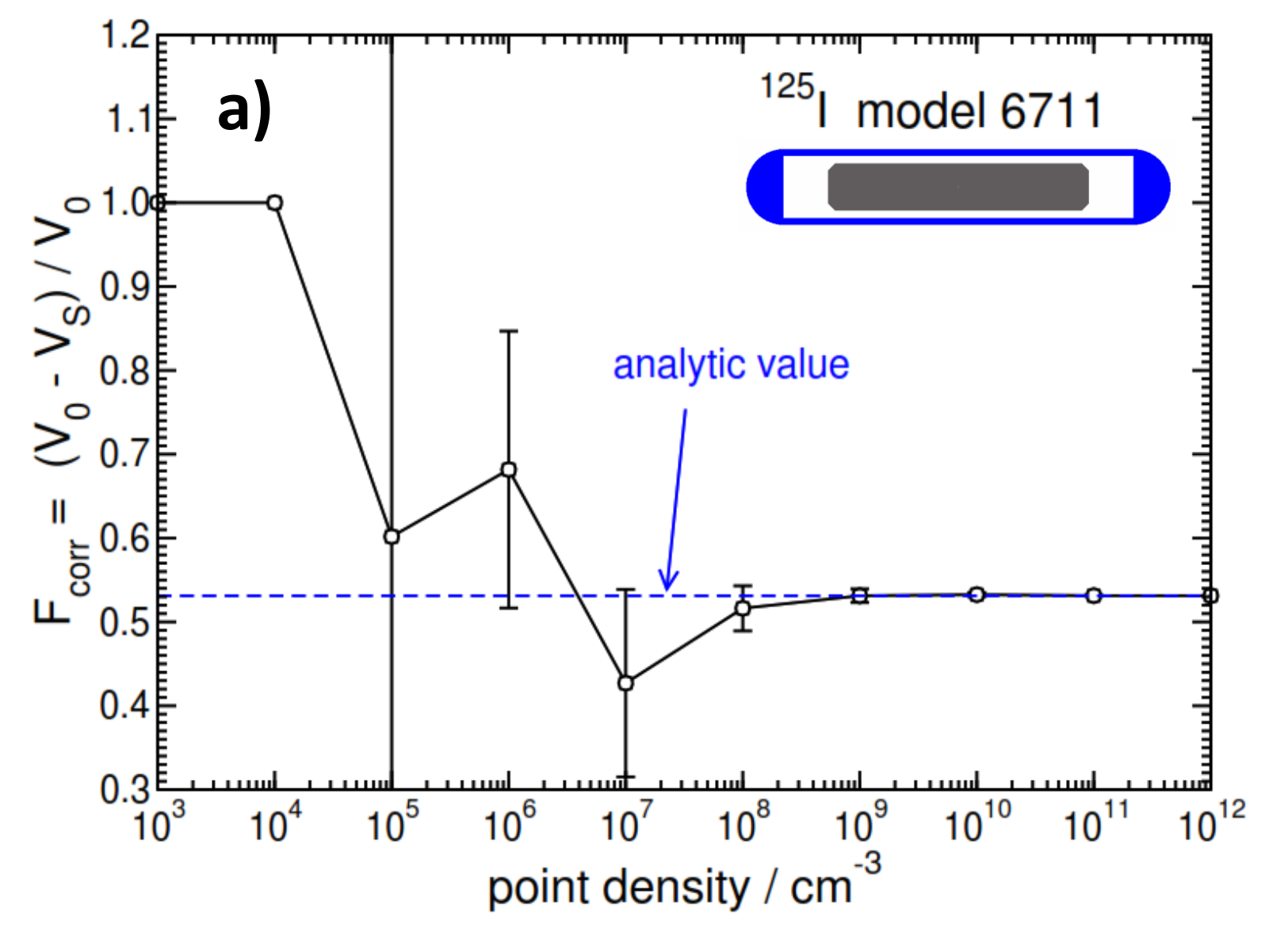}
\includegraphics[scale=0.32]{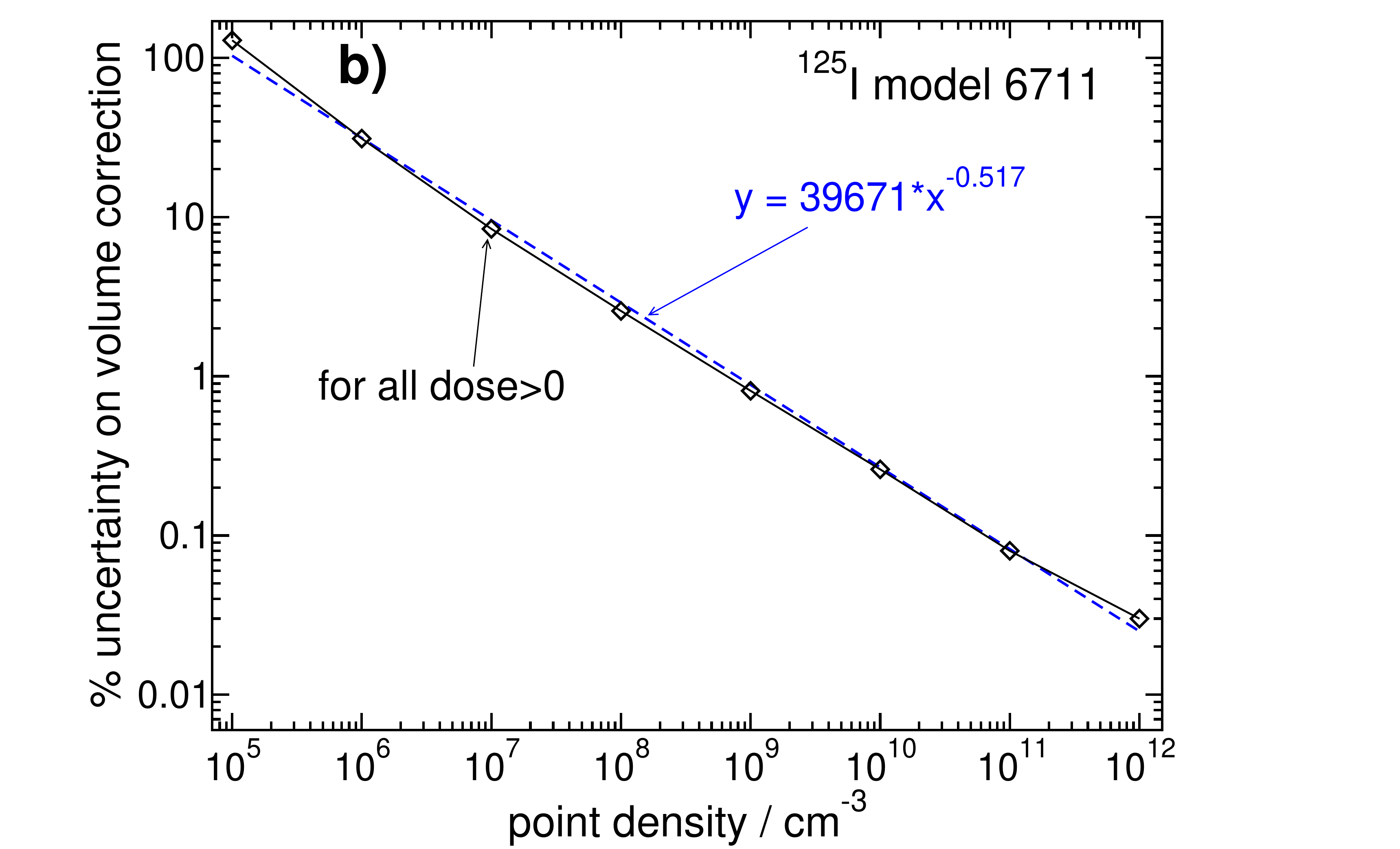}
\includegraphics[scale=0.42]{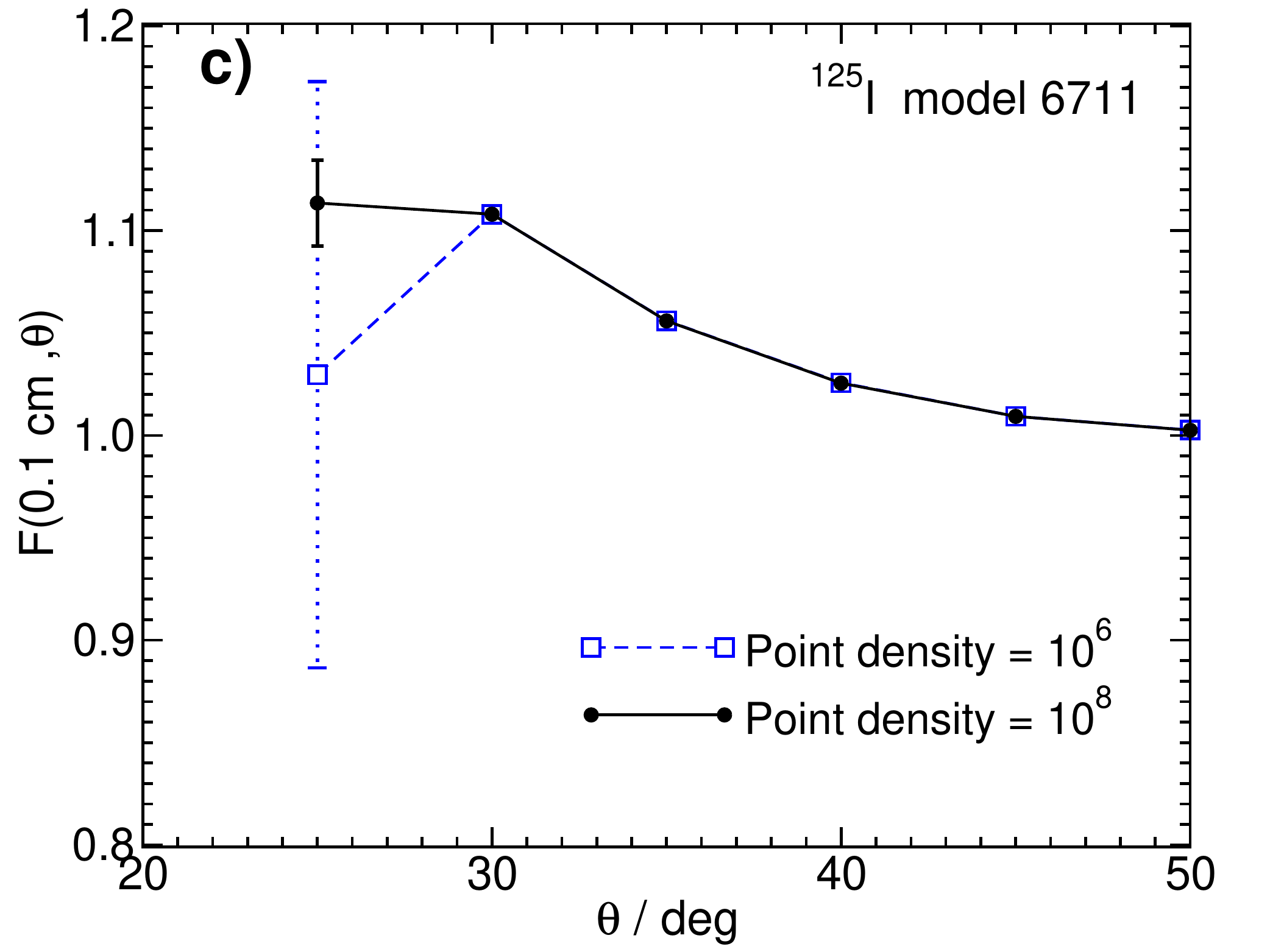} 
\captionl{(a) Calculated volume correction ($F_{\rm corr}$=($V_0$ $-$ $V_s$)/$V_0$) compared to the analytic volume correction as a function of
different point densities for a region very close to a source (\io model
6711). 
(b) Calculated average source volume correction uncertainty for all voxels
which overlap the source (with  98\% threshold coverage)
as a function of point density. (c) TG-43 anisotropy function calculated with
$10^6$ and $10^8$~cm$^{-3}$ point densities with uncertainties calculated
with \ebc.  Less than 1\% of the voxels require a source volume correction  
for this calculation.
\label{fig_Fcorr} }
     \end{center}
   \end{figure}

Figure~\ref{fig_Fcorr}a presents the source volume correction, $F_{\rm
corr} = (V_0-V_s)/V_0$, as a function of different point densities
($10^3-10^{12}~{\rm cm}^{-3}$)  after the above statistics are implemented
in \eb for a specific region very close to a 6711 source ($r=0.1$~cm,
$z=0.09$~cm). These values are compared to the analytic volume correction.
The error bars are statistical uncertainties generated by \ebc. When the
point density is very low ($10^3-10^4$ cm$^{-3}$ with a seed volume of
roughly 4$\times 10^{-3}$ cm$^3$), it is possible that no random points
fall in $V_s$ (\ie $N_s$=0) and hence no volume correction is made. When
the point density is increased ($10^5-10^7$ cm$^{-3}$), random points are
generated in both parts of the voxel. As point density continues to
increase, the code eventually calculates the correct phantom volume
compared to the analytic calculation for a point density of $10^{10}
~(10^{11})$ within 0.2\% (0.08\%) where the calculated uncertainty is 0.5\%
(0.09\%).

\FloatBarrier

Figure~\ref{fig_Fcorr}b shows, as a function of point density, the average uncertainty on the
volume correction for all voxels which overlap with a typical brachytherapy
source ($^{125}$I model 6711), have doses $>$0 (\ie are not completely
covered by the source), and have more than 2\% of the voxel volume outside
of the source.  This last condition is to avoid the very occasional extreme
fluctuation in uncertainty for these very small regions outside the source.
As the point density is increased, the dose uncertainties decrease
proportional to $1/\sqrt{PD}$ as seen in equation~\ref{eq_vol_only}.

  Figure~\ref{fig_Fcorr}c presents the calculated TG-43 anisotropy function 
for two different point densities ($10^6$ vs $10^8$). 
Values are the same for cases in which volume corrections are not
needed, but differ (albeit within statistical uncertainties) for
regions close to the source (\eg r= 0.1~cm, $z =0.09$ cm or $\theta=25^{\circ}$), where the voxel
overlaps with the source. Such large differences explain why there are
often large differences outside statistics between CLRPv1 (\BDc) and CLRPv2
(\ebc) results for voxels overlapping the source since the CLRPv1 data did
not include the uncertainties on the volume correction. Fortunately,
since the CLRPv1 data were only wrong in very high-dose regions or very
small regions close to
the source, they would have no practical effects on clinical dose
evaluations using these data.

When the point density is increased 10 times (e.g., from $10^8$ to $10^9$),
the volume correction calculation time increases approximately tenfold
(e.g., from 0.09 to 0.93 sec for a single seed case). For the database
calculations a point density of $10^{10}$~cm$^{-3}$ was used which implies
sub-1\% average uncertainty on the volume corrections for voxels overlapping the
sources.

\FloatBarrier

\subsection{TG-43 dosimetry parameters calculations}\label{methods:tg43}
Following the 1995 TG-43 protocol~\cite{Na95,Ri04}, the 2D dose-rate
distribution around a sealed brachytherapy source is determined as:
  \eqn{\dot{D}(r,\theta) = S_K\cdot\Lambda\cdot\frac{G_L(r,\theta)}
      {G_L(r_0,\theta_0)}\cdot g_{L}(r)\cdot F(r,\theta), \label{eq_tg43}}
where $r$ is the distance from the center of the source's radioactivity 
to the point
of interest, $r_0$ is the reference point which is defined at (1~cm,
90$^{\circ}$) on the transverse source axis, and $\theta$ denotes the polar
angle determining the  point of interest. Other quantities and functions are
defined elsewhere in this paper, and are discussed in  more detail in the
TG-43 reports.

\begin{figure}[ht]
\begin{center}
 \includegraphics[scale=0.67]{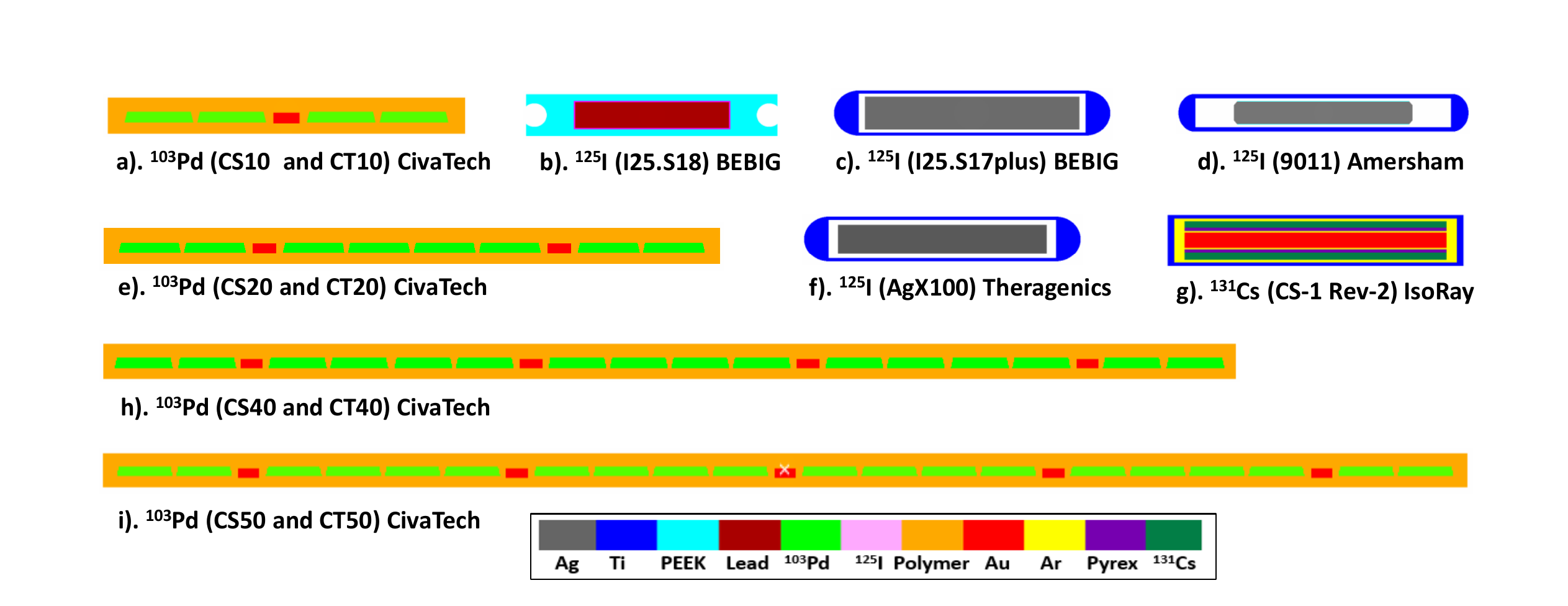}   
\captionl{Low-energy, LDR photon-emitting brachytherapy sources added to
CLRPv2. Source geometries are displayed using the {\tt egs\_view}
visualization tool of egs++. A consistent color coding format for different
materials is used throughout the database. 
Source dimensions are not to scale from one source to the next, but are to
scale for each source.
\label{fig_new_sources} }
\end{center}
\end{figure}

TG-43 dosimetry parameters are calculated for 40 low-energy LDR
photon-emitting brachytherapy sources. The 13 sources added
to the CLRPv2 database are shown in Figure~\ref{fig_new_sources}. 
A detailed description of all 40 sources is available online in the database.

Each source is modelled using the egs++ class library~\cite{Ka18} geometry
module. The egs++ geometry models of all sources will be distributed freely
with \ebc. For each source, the results of three 3D MC dose
calculations in full-scatter water phantoms with different voxel
resolutions (0.1, 0.5, and 1~mm$^3$) and the results of two MC 
simulations of air-kerma strength
measurements (Point and WAFAC detectors) are imported into an in-house
Python software tool, which extracts all TG-43 data. The same Python
suite was used to produce the CLRPv1 database.~\cite{TR08b,TR08c} Dose
values are tabulated as a function of distance from the source and polar
angle relative to the long (Z) axis of the source. Dose values are
interpolated bilinearly using the nearest neighboring voxels when the point
of interest does not fall at the center of a voxel.  The database specifies
the effective active
length of each source, $ L_{\rm eff}$, which accounts for the effect of the
physical shape of the radioactive material distribution inside the source.
It is calculated using the TG-43U1 effective line source
length approximation~\cite{Ri04,Ri05}, and in some complicated cases
(e.g.,$^{103}$Pd Best Medical model 2335) the maximum distance between
proximal and distal aspects of the radioactivity distribution is
used.~\cite{Ri05}

The radial dose function,
$g(r)$, is calculated using point {$(G_P(r,\theta))$}, and line
{$(G_L(r,\theta))$} source geometry functions, and is tabulated at
intervals of 0.01~cm for $0.05\leq r \leq0.1$~cm, 0.1~cm for
$0.1<r\leq1.0$~cm, and 0.5~cm for $5<r\leq10$~cm.
Values at $r=0.15$, 0.25,
and 0.75~cm are also reported. Fitting coefficients for the following
fifth-order polynomial functional form for $g_L(r)$ as proposed by Taylor and
Rogers~\cite{TR08} are also calculated. 
   \eqn{g_L(r) = \left(a_0 r^{-2} + a_1 r^{-1} + a_2 + a_3 r + a_4 r^2 + a_5
   r^3\right)e^{-a_6 r}.}

For each source's $g(r)$, the mean residual deviations from the actual data
for the best fit functions are also reported (all $\leq 0.2\%$).  The
closest radius included in the fit is specified and varies between 0.03 and
0.2~cm.  It varies since the very closest points sometimes ruin the entire
fit and/or are inside the source.

Results from extrapolating  $g_L(r)$ values at larger radial
distances from 10 to 15~cm using the above fitting coefficients agreed
well with MC results from \eb simulations in a larger water phantom (40~cm
height and 20~cm radius) for a representative source for each isotope (mean
differences in $g(r)$ values $\leq 0.7\%$ for $10 < {\rm r} \le 15$~cm).
Changes in $g_L(r\leq10$~cm)  and $F(r\leq10$~cm,~$\theta$) values  were
insignificant ($\leq 0.1\% $) when we performed calculation in the larger
40~cm phantom rather  than the standard 30~cm phantom.

The 2D anisotropy function, $F(r,\theta)$, describes the variation in dose
distribution in polar angle due to photon scattering and attenuation, as
well as self-absorption and oblique filtration in source encapsulation.
$F(r,\theta)$ values are calculated using the line source approximation and
tabulated at radii of 0.1, 0.15, 0.25, 0.5, 0.75, 1, 2, 3, 4, 5, 7.5
and 10~cm and at 32 polar angles with a minimum resolution of $5^{\circ}$.
For some sources (e.g., \pd CivaThin sources, and \io model 9011) with
smaller diameters ($\sim0.03~$cm) than conventional brachytherapy sources,
values at radii such as 0.05 or 0.07~cm are included.

The 1D anisotropy function (anisotropy factor), $\phi_{an}(r)$~\cite{Ri04},
is defined as the ratio of the solid-angle-weighted dose-rate averaged over
4$\pi$ steradians, to the dose-rate at the same r distance on
the transverse plane. These values are calculated by integrating the solid angle weighted dose-rate over
$0^{\circ}$ $\leq \theta \leq$ $ 90^{\circ}$. The $\phi_{an}(r)$ factor is
used in treatment planning when the source orientation is being ignored.

\subsection{Data additional to TG-43 parameters}
\label{methods:additional data}

In addition to TG-43 parameters, the database provides along and away
dose-rate per air-kerma strength data (cGy h$^{-1}$ U$^{-1}$), 
mean photon energy ($\bar{E}_\gamma$; keV) for all
LDR sources, as well as primary and scatter-separated (PSS) dose data for
several sources.

Along and away dose-rate data in Cartesian coordinate are normalized to $S_K$,
air-kerma strength, and are tabulated at 12 away distances from 0 to
10~cm and 13 along points from 0 to 10~cm. 
The along and away tables can be used as part of the needed quality
assurance checks while commissioning a specific source in a brachytherapy 
treatment planning system~\cite{Ri18}.

PSS data~\cite{Ru05} are provided for the following representative sources:
 \pd TheraSeed 200, \io 6711 and 130.002 models, and Isoray \cs Rev2 model.
The data are tabulated at 24 radii from 0.10 to 10~cm
and 24 polar angles with a minimum resolution of $5^{\circ}$. High
resolution ($\Delta$ r = 1~mm, $\Delta\theta=1^{\circ}$) PSS data are also
provided. For the purposes of these calculations, scatter within the source
is not considered as scatter so that any photon escaping the source
encapsulation is considered a primary.  Doses are normalized to the total
photon energy escaping the encapsulation.  

Photon energy spectra for each LDR source model are calculated using
\ebc's surface count scoring option.  These were calculated in a water
phantom with the \eb spectrum-scoring routines ignoring photons scattered
back across the surface from the water and so effectively are the spectra
in vacuum.  Although these back-scattered photons do have a small effect on
the spectrum they have no effect on the spectrum's mean energy.  The
mean photon energy, $\bar{E}_\gamma$, for each source is in column 1 of
Tables~\ref{table:drc_Pd} and ~\ref{table:drc_I}. The statistical
uncertainty on the mean energy is less than 0.01\%.  Representative photon
energy spectra for the \pd TheraSeed 200, \io OncoSeed 6711, and \cs
Cs-1 Rev2 sources are presented on the main page of the CLRPv2 database.

\FloatBarrier

\subsection{Data validation}\label{methods:validation}
\label{data_validation}

\begin{figure}[ht]
   \begin{center}
\includegraphics[scale=0.75]{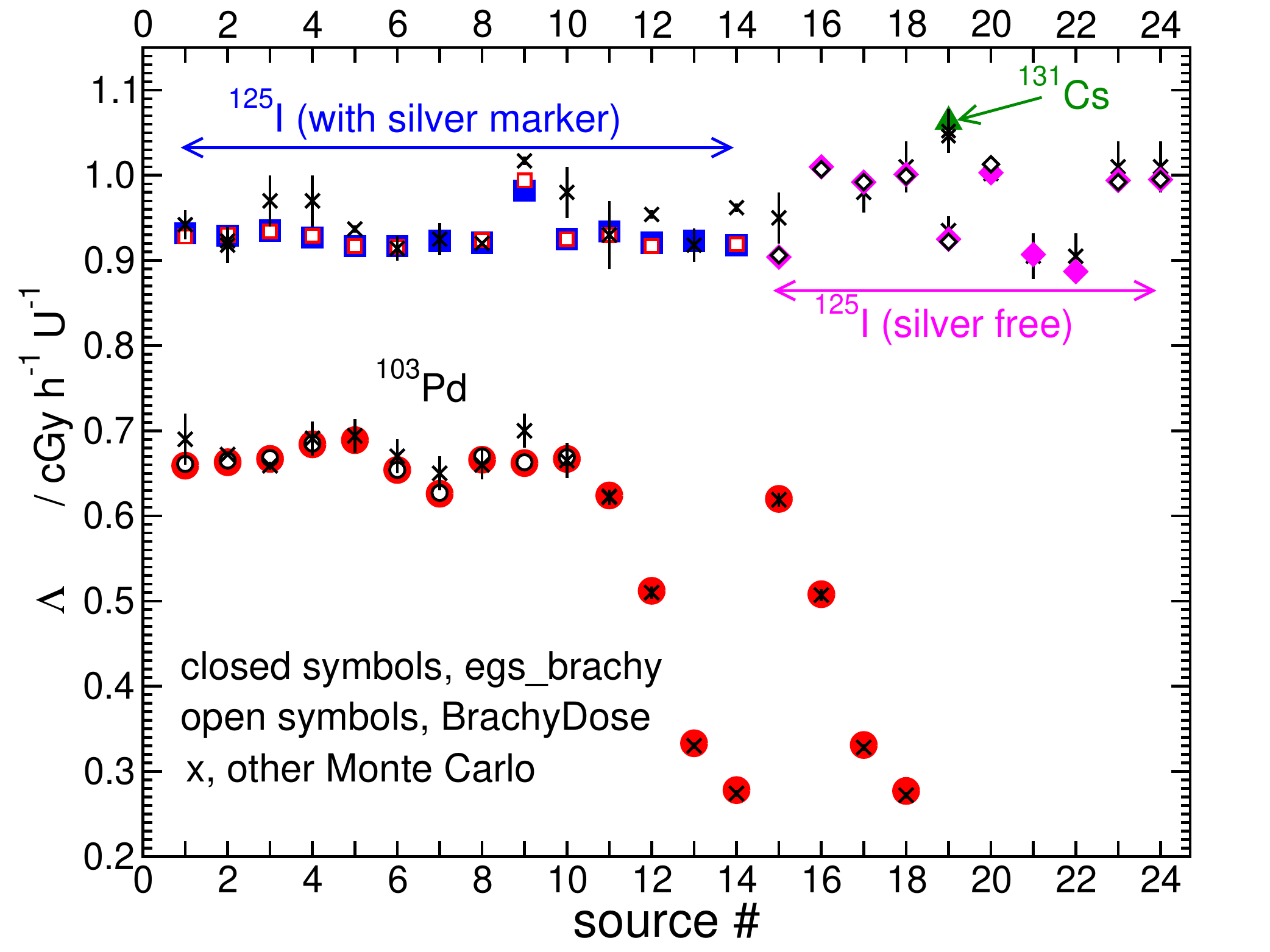}

\captionl{Values of DRC for different sources separated by radionuclide: 
\pd (red circles); \io without silver marker (blue squares) or
with silver marker (pink diamonds); and \cs (green triangle). CLRPv2 values
calculated with
\eb are closed symbols which can appear open if the \BD values overlap
almost exactly. Values calculated with \BD  are open symbols from
refs.~\cite{RR13, RR14,TR08b}. DRCs from other MC codes with
their uncertainty are also shown (x symbol). The statistical uncertainties
on \eb and \BD values are smaller than their symbols. The x-axis source
numbers are in Tables~\ref{table:drc_Pd} and \ref{table:drc_I}.
\label{fig_DRCcompare} }
     \end{center}
   \end{figure}

Data are validated by comparisons to previously published measurements and
calculations with other MC codes. DRC values are compared primarily to \BD
results from Rodriguez and Rogers~\cite{RR14,RR13} or from Taylor and
Rogers~\cite{TR08b}, as well as values from other codes and sources in the
literature - see Tables~\ref{table:drc_Pd} and~\ref{table:drc_I}. These
comparisons give percent differences as:
  \begin{equation}
   \%\Delta(\Lambda_1,\Lambda_2) = \frac{\Lambda_1-\Lambda_2}{\Lambda_1} 
   \times 100\%.
   \label{eq_Delta}
  \end{equation}
DRC values separated by radionuclide are shown in
Figure~\ref{fig_DRCcompare}. Due to the large amount of data, detailed
comparisons of $g(r)$ and $F(r,\theta)$ are omitted here, but the
comparisons are available in the database along with further DRC
comparisons when available.

In general, the CLRPv2 DRC values show excellent agreement with the \BD
results which modelled the same geometry with a maximum
difference of 0.6\% from the data of Rodriguez and Rogers~\cite{RR14}
and/or Taylor and Rogers~\cite{TR08b}. This excludes the differences due to
changes in how the source geometry was modelled as discussed below.  The
average difference is $(+0.17\pm0.26)$\% for \io sources and $(-0.24\pm0.17)$\%
for \pd sources which is excellent agreement given the statistical
uncertainty of 0.3\% on the earlier \BD results.

The mean DRC value for most \pd sources is 0.664 (cGy~h$^{-1}$~U$^{-1}$)
with a sample standard deviation of 0.017. The exceptions are the 
plastic encapsulated CivaTech source families of CivaString (CS10--CS50)
and CivaThin (CT10--CT50) with values ranging from 0.624 to 0.277
(Table~\ref{table:drc_Pd} and Figure~\ref{fig_DRCcompare}).
As the source length is increased from 1~cm (CS10 or CT10 sources) to 5~cm
(CS50 or CT50 sources), the DRC values decrease. This is due to the
increase in the average photon pathlength from the source to the reference
point on the axis at 1~cm and hence the increased attenuation in the water
phantom.

The corresponding  mean DRC value for all \io sources is 0.943 (cGy h$^{-1}$
U$^{-1}$) with a sample standard deviation of 0.037 (Table~\ref{table:drc_I}). The DRC value for the \cs Rev2 source is 1.06 and larger than
the average values for \pd and \io due its to relatively higher-energy
spectrum, which leads to less photon attenuation in the water phantom.
The mean photon energies calculated for all \pac, \ioc, and \cs sources are
20.55, 27.68, and 30.29~keV respectively (Tables~\ref{table:drc_Pd} and
~\ref{table:drc_I}, column 1).

For all \io source models containing a silver marker (indicated by a
superscript `a' in Table~\ref{table:drc_I}), the average DRC is 0.929
cGy~h$^{-1}$~U$^{-1}$ which is about 3.5\% lower than the average DRC of
0.962~cGy~h$^{-1}$~U$^{-1}$ for sources without a silver marker.  This is
due to the lower-energy (22, 25~keV) characteristic x rays from
silver~\cite{Cy90,RR13,RR14,NC07,He01} which reduce the average energy of
the spectrum from the source. The mean energy of \io sources with a silver
marker is 27.35~keV which is 0.8~keV lower than the average energy of
28.14~keV from sources without a silver marker (Table~\ref{table:drc_I},
column 1). The IPlant 3500 with a silver marker (source 9) is an outlier,
with a higher DRC value of 0.982~cGy~h$^{-1}$~U$^{-1}$.  This is because
the radioactive material is not directly distributed on the silver marker's
surface~\cite{Ri02}, leading to a higher mean photon energy and DRC value.
The silver marker in \pd sources has virtually no effect since the initial
\pd photons do not excite the silver x rays.
In Figure~\ref{fig_DRCcompare}, those \io sources  with DRCs $\geq0.99$ are
silver-free but not all silver-free sources have higher DRCs (\ie sources 15,
19, 21 and 22). The BT-125-1 model (source 15) has a palladium marker  and
the LS-1 model (source 19) has a platinum/iridium marker. These high-Z markers
mainly generate x-ray spectral components similar to sources containing a
silver marker~\cite{ He01}. The spectra of S18 sources~\cite{Ab10c}
(sources 21 and 22) have a small number of low-energy x rays (10 to 15~keV)
due to the composition of their lead-glass marker. There are not enough
low-energy x rays to significantly decrease the average energy of the
spectrum. However they do increase the air-kerma strength (due to the large
mass energy absorption coefficient of air at these lower energies) but do
not affect the dose at the reference point since they are attenuated in the
water. Together these effects result in a lower DRC value.

There are some sources that show a significant DRC discrepancy when
comparing \eb and earlier \BD values\cite{TR08b,RR13,RR14}.  These are
denoted by a superscript 'f' in Tables~\ref{table:drc_Pd}
and~\ref{table:drc_I}.  For these four sources, the original \BD air-kerma
calculations mistakenly included a thin cylindrical shell of water around
the sources which decreased the air-kerma strength and hence increased the
DRC values. After the source geometries were fixed the mean DRC value
deviations between \BD and \eb values is reduced to~$<$0.1\% with
differences ranging from -0.6\% to 0.4\%. A source with different geometry
models in \eb and \BD (source 13, Table~\ref{table:drc_I}) is omitted from
these differences.

Sources for which no \BD results are available have  DRCs which are within
the statistical uncertainties of other published values.

\begin{table} [h]
\begin{center}
\caption{\sf Dose-rate constant values for \pd and \cs sources
calculated using \eb (`This work'), \BD (`RR' and `TR')\cite{RR14,TR08b}, other codes (`Other MC'), and TG-43 consensus data\cite{Ri04,Ri07a,Ri17} (TG43$_{\rm con}$).  The \eb and \BD DRC values shown are for the NIST WAFAC geometry.  Statistical uncertainties are $\le$0.3\% (BD), $\le$0.02\% (eb), and otherwise shown in brackets (uncertainty in the last digit). Percent differences are given between results for \eb and the best \BD geometry [$\%\Delta$(eb,BD)], as well as with  TG43$_{\rm con}$ \newline { [$\%\Delta$ (eb,TG43$_{\rm con}$)]}.
The mean emitted photon energy ($\bar{E}_\gamma$)
of each source determined by \eb is indicated.\vspace{-5mm}\\
\label{table:drc_Pd}}
\begin{small}
\begin{tabular}{lccccccc}
\hline
& \multicolumn{6}{c}{Dose-rate constant $\Lambda$ (cGy h$^{-1}$ U$^{-1}$)}  \\
\cline{2-8}
\vspace{-0.5mm}{Source model} ($\bar{E}_\gamma$/keV)& This & \footnotesize{RR~\cite{RR14}} & \footnotesize{TR~\cite{TR08b}} & \footnotesize{$\%\Delta$} & \footnotesize{Other} &  \footnotesize{TG-43$_{\rm con}$}  & \footnotesize{$\%\Delta$}\\
&  work(eb) & \scriptsize{BD} & \scriptsize{BD} & \scriptsize{(eb,BD)} &
\scriptsize{MC} &    & \scriptsize (eb,TG43$_{\rm con}$)\\
\hline
 {\bf $^{103}$Pd}  	    &	   &	   	&    	&		&      &	\\
 1 IAPd-103A${\color{blue}^a}$ (20.51) 	&	0.659	&	0.661	&
0.687${\color{blue}^f}$	 &	-0.3	&	 0.69(3)~\cite{Me06}  &
0.693(31)~\cite{Ri17} & -5.2\\
 2 Med3633${\color{blue}^a}$ (20.50)	&	0.663${\color{blue}^b}$	&	0.665${\color{blue}^g}$	&	0.650	&	-0.3	&	 0.672~\cite{Ri01c}  & 0.688~\cite{Ri04}  & -3.8 \\
3 BT-103-3${\color{blue}^a}$ (20.50)     	&	0.667	&	0.668${\color{blue}^e}$	&	0.671	&	0.1	&	 0.659(5)~\cite{De02d} & n/a  & n/a \\
4 TheraSeed 200${\color{blue}^{c,i}}$ (20.51)	&	0.684 &	0.685${\color{blue}^g}$	&	0.694 	&	-0.1	&	 0.691(20)~\cite{MW02} & 0.686~\cite{Ri04} & -0.3 \\
5 TheraSeed 200${\color{blue}^{d,i}}$ (20.55) 	&	0.689	& n/a		&	n/a	&	n/a	&	 0.694(20)~\cite{MW02} & n/a & n/a  \\
6 BestPd-103 2335 (20.54)            &	0.654${\color{blue}^b}$	&	0.654	&	0.650	&	0.0	& 	 0.67(20)~\cite{Me01b} & 0.685~\cite{Ri07a}  & -4.7 \\
7 BrachySeed Pd-1 (20.55)        	&	0.626	&	0.627	&	0.632${\color{blue}^h}$	&	-0.2	&	 0.65(20)~\cite{Me03} & n/a  & n/a \\
8 BEBIG Pd-103 (20.50)         &	 0.666${\color{blue}^b}$  	&	0.670${\color{blue}^e}$	&	0.685${\color{blue}^f}$	&	-0.60	&	 0.660(17)~\cite{DW01} & n/a & n/a  \\
9 1031L${\color{blue}^i}$ (20.47)     	&	0.662${\color{blue}^b}$	&	0.663	&
0.663	&	-0.1	&	 0.70(20)~\cite{Me00b} &
0.701(20)~\cite{Ri17}  & -5.9 \\
10 1032P${\color{blue}^i}$ (20.54)  	&	0.667${\color{blue}^b}$	&	0.669${\color{blue}^j}$	&
0.669	&	-0.3	&	 0.665(21)~\cite{WH05} &
0.671(19)~\cite{Ri17} & -0.6 \\

11 CivaString CS10 (20.59)	& 0.624	&	n/a	    &	n/a  	&	n/a
&	 0.623(8)~\cite{Ri14} & 0.641(17)~\cite{Ri17} & -2.7 \\
                 	&   	&		    &	     	&	    	&	 0.622(9)~\cite{Re14}  \\
12 CivaString CS20 (20.58)	& 0.512	 &	n/a	    &	n/a  	&	n/a 	&	 0.510(7)~\cite{Ri14}  & n/a & n/a \\
13 CivaString CS40 (20.59)	& 0.333	 &	n/a	    &	n/a  	&	n/a 	&	 0.330(4)~\cite{Ri14} & n/a & n/a  \\
14 CivaString CS50	(20.58) & 0.279	 &	n/a	    &	n/a  	&	n/a 	&	 0.274(4)~\cite{Ri14}  & n/a & n/a \\	
15 CivaThin CT10 (20.58)	& 0.620	 &	n/a	    &	n/a  	&	n/a 	   &	 0.619(8)~\cite{Ri14}   & n/a & n/a  \\	
16 CivaThin CT20 (20.58)	& 0.509	 &	n/a	    &	n/a  	&	n/a     	&	 0.507(7)~\cite{Ri14}  & n/a & n/a \\	
17 CivaThin CT40 (20.58)	& 0.331	 &	n/a	    &	n/a  	&	n/a     	&	 0.328(4)~\cite{Ri14}  & n/a & n/a \\	
\vspace{2mm}18 CivaThin CT50 (20.58)	& 0.277	 &	n/a	    &	n/a  	&	n/a 	&	 0.272(4)~\cite{Ri14} & n/a & n/a \\	

{\bf $^{131}$Cs}  	&	 	&	 	&	 	&	 	&	 \\
19 CS-1 Rev2 (30.29)    	&	1.063	&	 n/a  	&	 n/a  	&
n/a  	&	 1.046(19)~\cite{Ri07} & 1.056(13)~\cite{Ri17} &	 0.7 \\
  &   	&		    &	   	&  	&  1.052(26)~\cite{ MR08} 	\\

\hline

\end{tabular}
\end{small}
\end{center}
\mbox{~}\vspace{-5mm}\\
$^a$ \scriptsize{Sources which include silver marker}\\
$^b$\scriptsize{ There is a change in the model of the source described in the database}\\
$^c$ \scriptsize{\pd Theragenics `Light seed': thickness of Pd radioactive
coating layer on graphite surface is 2.2~$\mu$m}\\
$^d$ \scriptsize{\pd Theragenics `Heavy seed': thickness of Pd radioactive 
coating layer on graphite surface is 10.5~$\mu$m}\\
$^e$ \scriptsize{\BD values recalculated for this work after correcting the geometry}\\ 
$^f$ \scriptsize{\BD values differ from \eb values because of water envelope geometry known to be the cause of change}\\
$^g$ \scriptsize{There were some geometry changes reported in RR\cite{RR14}
but the exact difference is unknown since the TR\cite{TR08b} geometry file
is lost}\\
$^h$ \scriptsize{There is an unexplained change since we do not have the MC input 
file of TR\cite{TR08b}}\\
$^i$ \scriptsize{Considering air and water in the central `empty' region of the source, 
there is a negligible difference (0.05\%) in DRC}\\ 
$^j$ \scriptsize{This value is taken from ref.~RR\cite{RR13} }\\ 
\end{table}

\typeout{***********end first table********************}
\typeout{***********end first table********************}
\typeout{***********end first table********************}

\clearpage

\begin{table}
\typeout{***********start table 2 ********************}
\begin{center}
\caption{\sf Same as Table~\ref{table:drc_Pd}, except for \io sources.    
        \label{table:drc_I}}
        \begin{small}
\begin{tabular}{lccccccc}
\hline
& \multicolumn{6}{c}{Dose-rate constant $\Lambda$ (cGy h$^{-1}$ U$^{-1}$)}  \\
\cline{2-8}
\vspace{-0.5mm}{Source model} ($\bar{E}_\gamma$/keV) & This & \footnotesize{RR~\cite{RR14}} & \footnotesize{TR~\cite{TR08b}} & \footnotesize{$\%\Delta$} & \footnotesize{Other} &  \footnotesize{TG-43$_{\rm con}$}  & \footnotesize{$\%\Delta$}\\
&  work(eb) & \scriptsize{BD} & \scriptsize{BD} & \scriptsize{(eb,BD)} &
\scriptsize{MC} &    & \scriptsize (eb,TG43$_{\rm con}$)\\
\hline
\hline
1 OncoSeed 6711${\color{blue}^a}$ (27.34)           	&	0.932${\color{blue}^b}$	&	0.928	&	0.924	&	0.4	&	 0.942(17)~\cite{Do06} & 0.965~\cite{Ri04}  & -3.5 \\
2 ThinSeed 9011${\color{blue}^a}$ (27.26)           	&	0.929	&
0.930	&	 n/a  	&  -0.1 &	 0.923(4)~\cite{Ke10b}  &
0.933(28)~\cite{Ri17}  &  -0.4 \\
3 EchoSeed 6733${\color{blue}^{a, c}}$ (27.24)	&	0.935	&	0.934	&	n/a	&	0.1	&	 0.97(3)~\cite{SM02a} & 0.980~\cite{Ri07a}  & -4.8\\
4 EchoSeed 6733${\color{blue}^{a,d}}$ (27.29)          	
&	0.927	&	n/a	&	0.929	&	-0.2	&	 0.97(3)~\cite{SM02a}  &	n/a \\
5 Braquibac${\color{blue}^a}$ (27.42)                	&	0.917	&	 n/a  	&	0.917	& 0.0  &    0.937(4)~\cite{Pi07} & n/a  & n/a\\
6 I25.S17${\color{blue}^a}$ (27.28)         	&	0.917	&	0.915	&
0.916	&	0.1	&	 0.914(14)~\cite{Ly05a} &
0.933(25)~\cite{Ri17} & -1.7 \\
7 I25.S17plus${\color{blue}^a}$ (27.30)   	&	0.923	&	 n/a  	&
n/a	&	 n/a  	&	 0.925(19)~\cite{Pa13} &
0.940(25)~\cite{Ri17} & -1.8\\
8 IS-12501${\color{blue}^a}$ (27.19)       	&	0.921${\color{blue}^b}$	&	0.924	&	0.924	&	-0.3	&	 0.92~\cite{Ib02} &  0.940~\cite{Ri04} & -2.1 \\
9 IPlant 3500${\color{blue}^{a,l}}$ (28.30)	&	 0.982${\color{blue}^b}$  	&	0.987${\color{blue}^e}$	&	0.994${\color{blue}^f}$	&	-0.5	&	 1.017(5)~\cite{Ri02} & 1.014~\cite{Ri07a} & -3.3 \\
10 IAI-125A${\color{blue}^a}$ (27.27)      	&	0.925${\color{blue}^b}$	&	0.925	&	0.925	&  0.0	&	 0.98(3)~\cite{Me02} & 0.981~\cite{Ri07a} & -6.1\\
11 125SL${\color{blue}^a}$ (27.17)      	&	0.934	&	0.931	&	0.93	&	0.3	&	 0.93(4)~\cite{Li02d} & 0.953~\cite{Ri07a} & -2.0\\
12 130.002${\color{blue}^a}$ (27.23)     	&	0.921	&	0.917	&
0.917	&	0.4	&	 0.954(5)~\cite{Ka01b} &
0.954(43)~\cite{Ri17} & -3.6\\
13 AgX100$^{\color{blue}^a}$ (27.29)          	&	0.923	&
0.900$^{\color{blue}^h}$   	&	 n/a  	&	 2.5       	&
0.918(20)~\cite{Mo12a} & 0.952(43)~\cite{Ri17} & -2.5\\
14 BT-125-2${\color{blue}^a}$ (27.35)      	&	0.918${\color{blue}^b}$ 	&	na	&	0.916	&	0.2	&	 0.962(5)~\cite{So02} & n/a 	&	 n/a\\
15 BT-125-1 (27.28)   	&	0.904${\color{blue}^b}$ 	&	0.906	&	0.901	&	-0.2	&	 0.95(3)~\cite{Po00} & n/a 	&	 n/a\\
16 OncoSeed 6702 (28.36)        	&	1.010	&	1.007	&	1.000	&	0.3	    &	 1.009~\cite{Ma98e} & 1.036~\cite{Ri04}  & -2.6 \\

17 STM1251 (28.44)                	&	0.992${\color{blue}^b}$	&
0.992${\color{blue}^g}$	&	1.012	&	0.0	&	 0.98(2.4)~\cite{KW01} & 1.018~\cite{Ri07a} & -2.6 \\
18 Best 2301 (28.39)              	&	1.001	&	0.999	&	0.998	&	0.2	&	 1.01(3)~\cite{SM02} &  1.018~\cite{Ri04} & -1.7\\
19 LS-1 (27.83)        	&	0.925	&	0.922	&	0.922	&	0.3	&	 0.935(17)~\cite{Wi02a} & 0.972~\cite{Ri07a} & -5.1\\
20 I25.S06 (.S16) (28.16)        	&	1.003${\color{blue}^b}$   	&	1.004${\color{blue}^e}$	&	1.011${\color{blue}^f}$	&	-0.1	&	 1.002~\cite{He00} & 1.012~\cite{Ri04} & -0.9\\
21 BEBIG I25.S18${\color{blue}^j}$ (28.17)        &	0.907	&	 n/a  	&
n/a	&	 n/a  	&	 0.905(27)~\cite{Ab10c} &
0.893(32)~\cite{Ri17} & 1.5\\
22 BEBIG I25.S18${\color{blue}^k}$  (28.12)       &	0.887	&	 n/a  	&	 n/a	&	 n/a  	&	 0.905(27)~\cite{Ab10c}  &	 n/a &	 n/a\\
23 1251L${\color{blue}^i}$ (28.32)      	&	0.994	&	0.991	&	0.992	&	0.2	&	 1.01(3)~\cite{Me02a} & 1.038~\cite{Ri07a} & -4.4\\

24 Med3631 (28.36)       	&	0.995	&	0.995${\color{blue}^h}$	&	0.978	&	0.0	&	 1.01(3)~\cite{Ri01b} & 1.036~\cite{Ri04} & -4.1 \\
\hline
\end{tabular}
\end{small}
\end{center}

$^a$ \scriptsize{Sources which include silver marker.}\\
$^b$\scriptsize{ There is a change in the model of the source described in the database}\\
$^c$\scriptsize{~\io coating layer in EchoSeed 6733 is assumed to have a
thickness of 10$~\mu$m on surface of silver marker}\\
$^d$\scriptsize{~\io coating layer in EchoSeed 6733 is assumed to have a
thickness of 2~$\mu$m on surface of silver marker}\\
$^e$ \scriptsize{\BD values recalculated for this work after correcting the geometry}\\ 
$^f$ \scriptsize{\BD values differ from \eb values due to water envelope geometry 
causing the change}\\ 
$^g$\scriptsize{There were some geometry changes in RR\cite{RR14} but the
original TR\cite{TR08b} geometry coding is lost.}\\
$^h$ \scriptsize{There is an unexplained change since we do not have the MC input file of RR~\cite{RR14}}\\ 
$^i$\scriptsize{ Considering air and water in hole portion of the source, there is a negligible difference (0.05\%) in DRC}\\
$^j$\scriptsize{~ with minimum available \%18 Lead-glass (PbO) as a marker
in Bebig I25.S18 source}\\
$^k$\scriptsize{~ with maximum available \%32 Lead-glass (PbO) as a marker
in Bebig I25.S18 source}\\
$^l$ \scriptsize{Radioactive material in IPlant3500 source is far away from
silver marker, which causes a higher mean energy and also DRC value}\\

\end{table}

\FloatBarrier
   
\begin{figure}[ht]
   \begin{center}
  \includegraphics[scale=0.75]{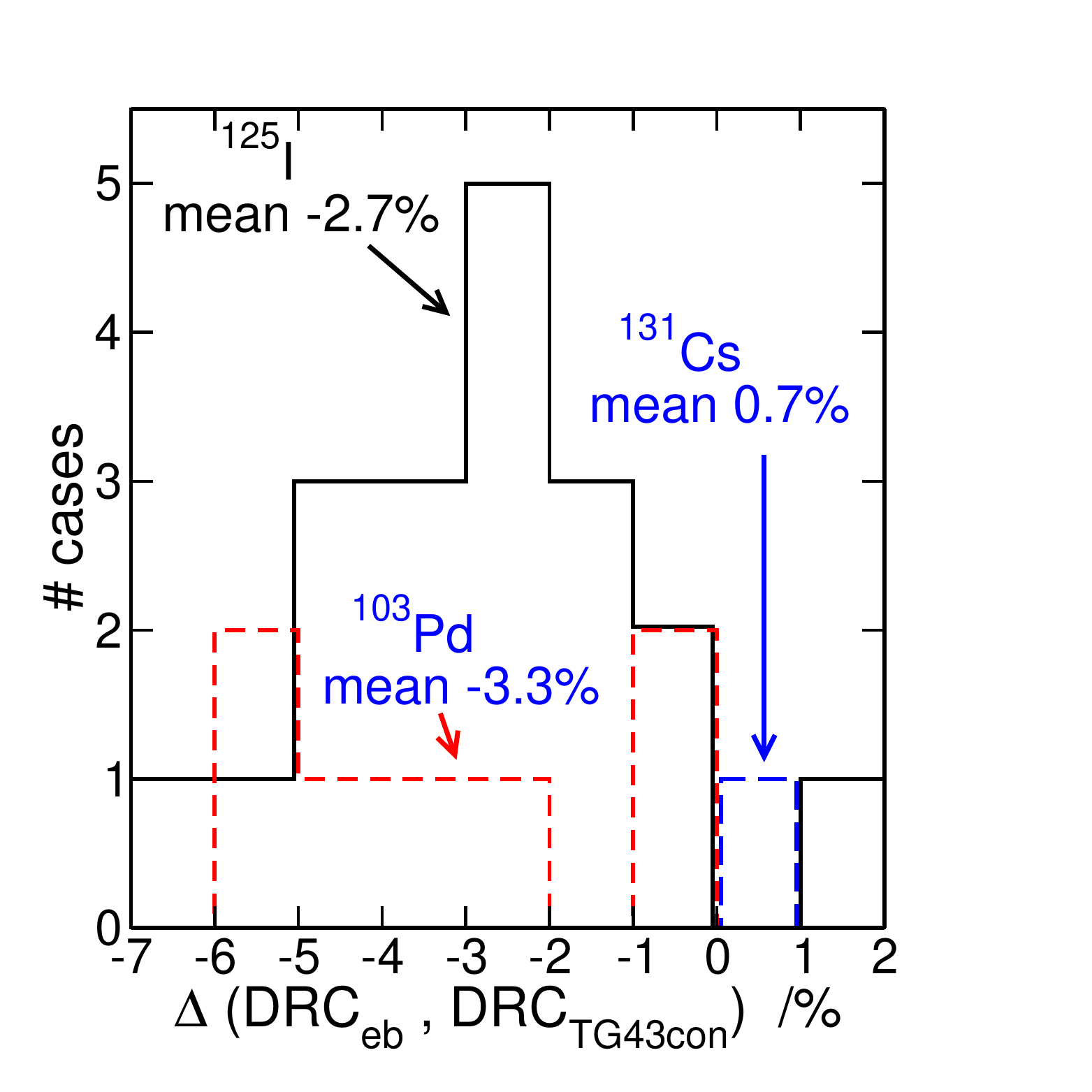}
  \captionl{ Distribution of DRC percentage differences between \eb and
  TG43 consensus values as defined by equation~\ref{eq_Delta} for \pd (red
  shorter dashed histogram), \io (black solid histogram), and \cs (blue
  longer dashed histogram) sources. Values are taken from last column in
  Tables~\ref{table:drc_Pd} and \ref{table:drc_I}.
  \label{fig_DRCdiff-eb&TG43} }
  \end{center}
  \end{figure}

As mentioned above, the TG-43 consensus DRC values for low-energy
brachytherapy sources have been determined by taking an average of the
average of MC calculations results and the average of the experimental
measurements.  Figure~\ref{fig_DRCdiff-eb&TG43} plots the data listed in
the last column in Tables~\ref{table:drc_Pd} and~\ref{table:drc_I}. It
shows that the Monte Carlo DRC values are generally lower than consensus
values. This is primarily  due to errors in the measured DRCs caused by
ignoring the intrinsic energy dependence of TLD detectors and the generally
large systematic uncertainties in the measurements\cite{RR14}. The mean
(range) differences between CLRPv2 and TG-43$_{\rm con}$ DRC values  for
\pac, \ioc, and \cs sources are -3.3\% (-6\%, -0.3\%), -2.7\% (-6.1\%,
1.5\%), and 0.7\%, respectively. As suggested before\cite{RR14}, this
indicates there is a need to revise many of the consensus values of DRCs
for LDR sources and possibly to base these solely on the Monte Carlo
calculated values. \vspace{5mm}

\FloatBarrier

 In general, $g_L(r)$ values are in agreement with \BD results within
statistical uncertainties for all sources studied, except for some regions
very close to the source where differences outside statistical
uncertainties and up to 3\% were found. This is because the statistical
uncertainties in the source volume correction needed for small $r$ values
 were not accounted for in the \BD calculations.

\begin{figure}[ht]
\begin{center}
\includegraphics[scale=0.5]{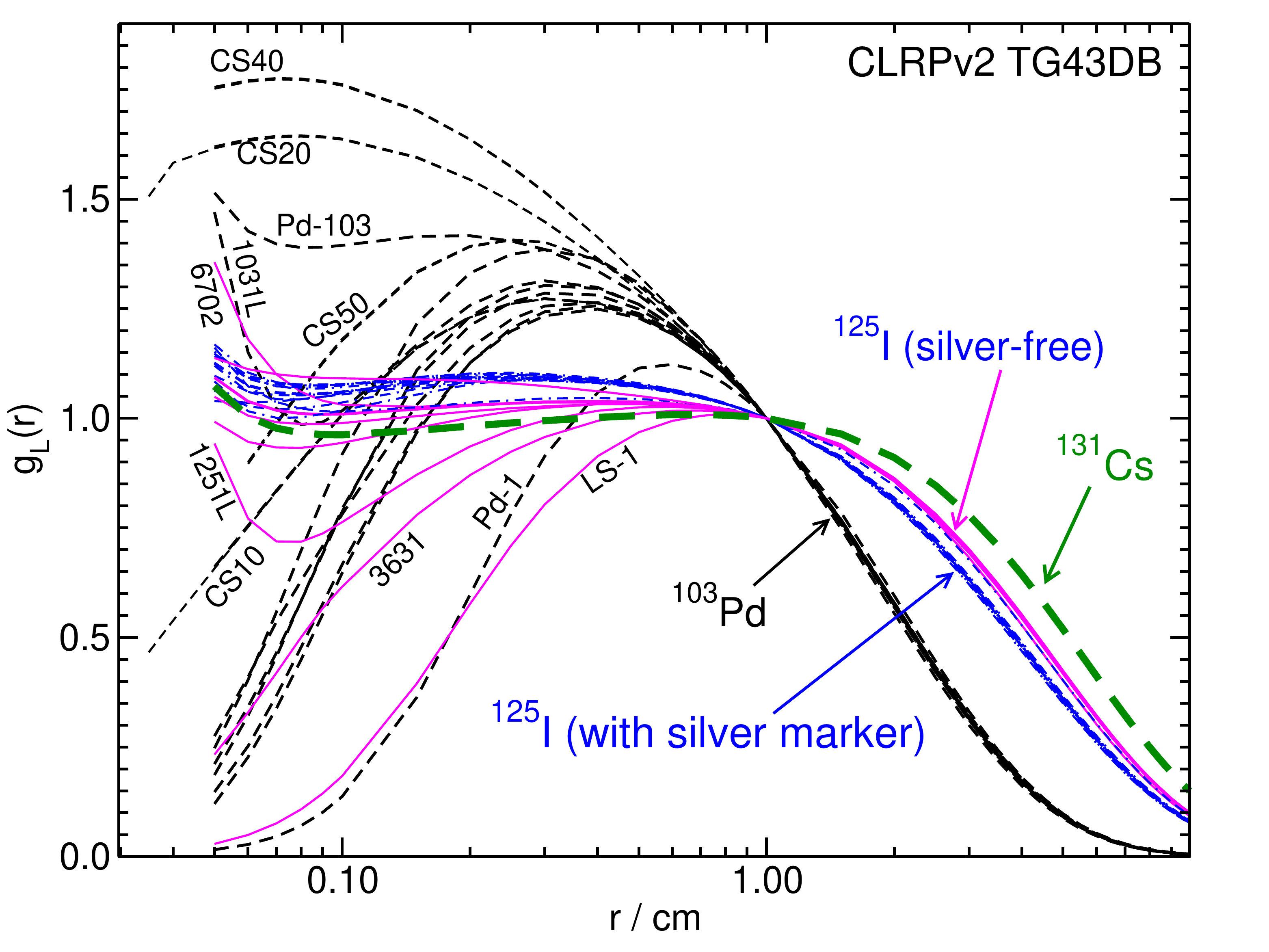}
\captionl{Radial dose function, $g_L(r)$, of all 40 sources including 17
\pd (black dashed lines), 22 \io (blue dashed lines with silver marker and
purple solid line without silver), and 1 \cs (green thick dashed line)
sources for CLRPv2 database as a function of distance and radionuclide. 
The $g_L(r\leq 1$~cm) values show quite different behaviours with a dose
build-up region for class A sources (e.g., \pd CS(CT)10, CS(CT)50, Pd-1,
and \io LS-1, 3631), and a flatter region with an upturn for the smallest
$r$ for class B sources (e.g., \pd CS(CT)20, CS(CT)40, Pd-103, and all \io
with a silver marker).  \label{fig_gofr}} 
\end{center} 
\end{figure}

Figure~\ref{fig_gofr} presents $g_L(r)$ values as a function of $r$ for all
the sources.  For sources of the same radionuclide, $g_L(r>$1~cm) values
are nearly indistinguishable except for differences between silver-marker 
and silver-free \io sources, for the same reasons outlined in
section~\ref{data_validation}.

For $ r \leq1$~cm, as discussed previously\cite{TR08}  the behaviour is
quite different depending on whether they are a class A source (two or more
sources of radiation separated by a radio-opaque marker) or a class B
source (radioactive material distributed along the length of the source
including at the center).  For class A sources, there is a dose
build-up along the transverse source axis (\pd CS(CT)10, CS(CT)50,
Pd-1, and \io LS-1, 3631 sources in Figure~\ref{fig_gofr}) because there is no radioactivity  on the
transverse axis. This, in turn, causes the dose to be low very
close to the source. For class B
sources (\pd CS(CT)20, CS(CT)40, Pd-103, and all \io sources with a
silver marker in Figure~\ref{fig_gofr}) $g_L(r<$1~cm) values are roughly
flat, with a slight upturn at the smallest $r$ values, except for the much
longer CivaTech sources. In class B sources, the radioactivity is uniformly
distributed and very close to the source. In this case the simple
geometry factor, $G_L(r,\theta)$, breaks down and only the radioactivity
directly on the axis matters, thus attenuation
in the walls of the source has a relatively smaller impact. 

\FloatBarrier
 
Plots in the database of $F(r,\theta)$ for each source show that most 2D
anisotropy function values agree within statistical uncertainties with
those calculated using \BDc. There are a few exceptions. For all CLRPv2 LDR
sources, the anisotropy functions for regions very close to the source show
differences vs CLRPv1, up to 46\% due to unspecified and
presumably large statistical uncertainties in the source volume correction
used with CLRPv1. The CLRPv2 values include this uncertainty and are thus
more reliable (see the effects in Figure~\ref{fig_Fcorr}c). There are other differences which are traceable to changes in the
geometric models of the seeds and these changes are specified in the database.

\begin{figure}[ht]
\begin{center}
\includegraphics[scale=0.5]{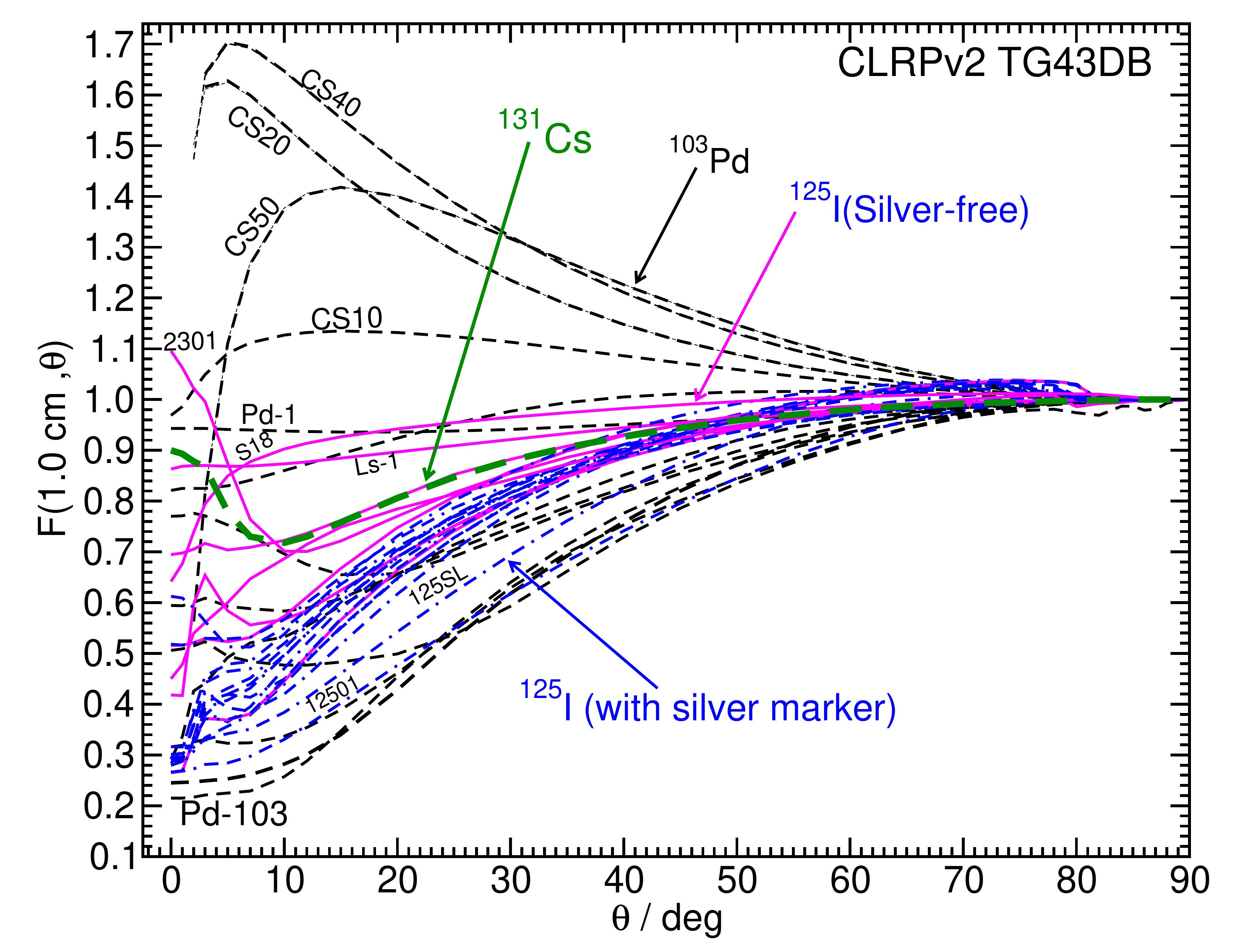}
\captionl{The 2D anisotropy function, $F(1.0{\rm~cm},\theta)$, of all 40
sources (17 \pac, 22 \ioc, and 1 \cs) in the CLRPv2 database as a function of
radionuclides and angle (degrees). The $F(1.0{\rm~cm},\theta \leq 20^o )$
variations for \pd sources (black dashed lines) are more significant
than those for \io sources
with silver marker (blue dash-dot lines) or without silver marker (purple
solid lines), and the \cs source (green thick dashed line).
\label{fig_Frtheta_all} }
     \end{center}
   \end{figure}

Figure~\ref{fig_Frtheta_all} presents $F(1.0,\theta)$ values for all the
sources in the database. Generally it is seen that the end caps attenuate
the doses near 0 degrees. Sources with beads at either end separated by a
marker (\eg Pd-1 and Ls-1 sources) maintain anisotropy values close to
unity even close to 0 degrees. The long CivaTech sources can have very
high values at 5 to 10 degrees because those points are close to the
radioactivity in the longer sources.\\

\clearpage

\begin{figure}[ht]
\begin{center}
\includegraphics[scale=0.47]{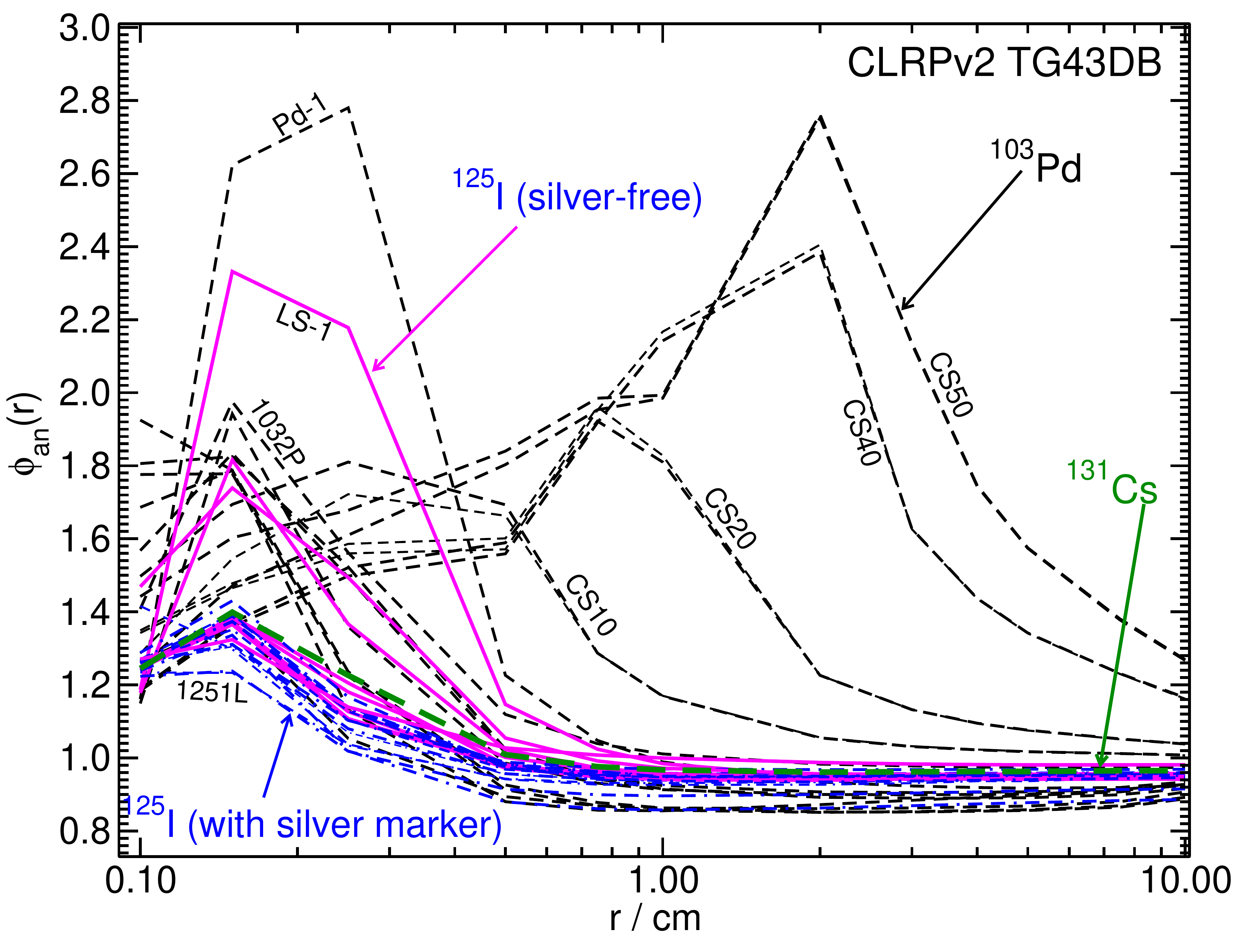}
  \captionl{The 1D anisotropy function, $\phi_{an}(r)$,   of all 40 sources
  including 17 \pa (black dashed lines), 22 \io (blue dash-dot lines with
  silver marker and purple solid lines without silver), and 1 \cs (green
  thick dashed line), as a function of distance.
  \label{fig_1Danisotropy_all} }
\end{center}
\end{figure}

\FloatBarrier
Some brachytherapy treatment planning systems use a 1D dosimetry formalism
by replacing the 2D anisotropy function in Equation~\ref{eq_tg43}, 
F($r,\theta$), with a 1D anisotropy factor, $\phi_{an}(r)$\cite{Ri04}.
Values of $\phi_{an}(r)$ for each source  are included in the database and
are shown in Figure~\ref{fig_1Danisotropy_all}.
Values of $\phi_{an}(r)$ varied between 0.85 and 1.0 for $r>1$~cm,
except for the longer CivaTech sources where the low value of the DRC, and
hence dose at the reference position (1~cm, 90$^{\circ}$), led to higher values
at larger distances due to source length. \\

\clearpage

\section{Data Format and Access}\label{format}

The CLRPv2 TG-43 parameter database 
is available online at
\url{https://physics.carleton.ca/clrp/egs_brachy/seed_database_v2} 
with a digital object identifier (DOI) at
\url{http://doi.org/10.22215/clrp/tg43v2}.
The website is hosted by Carleton University, in Ottawa, Canada. The main page
of the database lists all sources for which datasets are available,
information on the radionuclide source spectra and the half-lives used,
and information about spectrum average energies.

The CLRPv2 TG-43 database for low-energy LDR
brachytherapy sources contains roughly 315 data tables and 235 figures which
include approximately 65,000 data points. These datasets
include available and discontinued source models (for retrospective analysis).

For each source model, the following information is available in the CLRPv2
database:
\begin{itemize}  
\item A to-scale image showing a longitudinal cross-section of the source
model created using an egs\_view image of the actual egs++ model of the
source.

\item A description of the source's geometric model according to the
literature and manufacturer information, as implemented in
egs++. These models will be released for use with the open-source \eb application.

\item A table of DRC values, with absolute uncertainties, both in units of
cGy~h${}^{-1}$~U${}^{-1}$, along with values calculated with
\BD\cite{RR13,RR14,TR08b}, as well as other measured and calculated values
from different publications.

\item A figure comparing $g_L(r)$ data with corresponding values from Taylor and
Rogers~\cite{TR08b,TR08c} and from other papers in the literature.

\item A table of fitting coefficients for $g_L(r)$ using the Taylor and
Rogers modified polynomial function~\cite{TR08}. Extrapolating the $g_L(r)$
values using the current fitting coefficients was validated for larger
distances (10~cm$\leq r\leq15$~cm) with average differences of $<0.7$\%
from values based on MC simulations in a larger phantom (40~cm height and
diameter) .

\item A figure comparing $F(r,\theta)$ data at various values of $r$ with
data from Taylor and Rogers~\cite{TR08b,TR08c} and from other papers in
the literature.

\item Tabulated along-away dose data
(normalized to the air-kerma strength).

\item Low resolution PSS (primary-scatter separated) tabulated data for \pd TheraSeed 200,
\io 6711 and 130.002, and \cs Rev2 models, along with plots of
representative PSS data for these sources at different angles and radial distances.
High resolution PSS data files for these sources are also provided in CSV 
(comma-separated values) format.

\item A complete set of tabulated data for each individual source (DRC,
$g(r)$ and fitting coefficients, $F(r,\theta)$, and along and away tables)
in Microsoft Excel .xlsx format.

\end{itemize}

\section{Potential Impact}
\label{impact} 
Despite growing momentum towards adoption of model-based dose calculation
algorithms (MBDCAs)~\cite{tg186}, clinical brachytherapy dose prescriptions remain,
for the time being, based on the AAPM TG-43 dose calculation
formalism~\cite{Na95,Ri04,Ri04a,Ri07a,Ri17,Ri18,Pe12}. Given that the vast
majority of the TG-43 DRC consensus values are underestimated (on average
by 3\%), the more accurate MC-calculated datasets may, in future, be used
directly.  The CLRPv2 data may be used for research related to TG-43
dosimetry and brachytherapy planning, and may be considered in future
updates to the TG-43 consensus data. Additional brachytherapy sources may
be added to the database in the future, as new source models are released.
A few typos and inaccuracies in the source geometry descriptions on the
CLRPv1 database have also been corrected in the v2 database. Those
corrections are noted in the text on the database.  Finally, the complete
brachytherapy source geometry models described in the CLRPv2 database will
be released for use with the \eb EGSnrc application.  Release of all
benchmarked source models will enable \ebc's use for patient-specific
model-based dose calculations, thus supporting further research and
clinical adoption of MBDCAs as recommended by AAPM-ESTRO-ABG
TG-186~\cite{tg186}.

\section{Conclusion}
This database provides an update to the CLRP TG-43 dosimetry parameters
(CLRPv2) using the EGSnrc  \eb application for 40 low-energy
photon-emitting LDR brachytherapy sources (22 \ioc, 17 \pac, and 1 \cs), as well as source models to be distributed with \ebc.  Overall, the results are in good agreement with the previous CLRP TG-43 database (CLRPv1) which used the \BD
application to extract the TG-43 parameters for 18 \io and 9 \pd sources.
Statistical uncertainties, source volume corrections, and
modelling of several source geometries are all improved in comparison to CLRPv1
calculations. The 22 \io sources are categorized into two groups: with and
without silver marker. The general trends show clearly that \io
sources with silver markers have lower DRC values than those without. This is due to the contribution of lower-energy silver x rays which
reduce the mean energy of the source spectrum. With a silver marker, the
mean DRC value is 0.929 with a sample deviation of $\pm$
0.016 cGy~h${}^{-1}$~U${}^{-1}$ and for silver-free \io sources the mean
DRC is 0.962 $\pm$ 0.049 cGy~h${}^{-1}$~U${}^{-1}$.  For \pd sources, the
mean DRC value is 0.664 $\pm$ 0.017 cGy~h${}^{-1}$~U${}^{-1}$  except for
the plastic encapsulated CivaString and CivaThin source families. The DRC
value for the \cs source is larger (1.06~cGy~h${}^{-1}$~U${}^{-1}$) due to
the relatively high-energy spectrum compared to the other two
radionuclides. More details of other TG-43 parameters are extensively
presented in the CLRPv2 database which is hosted at Carleton University
website
(\url{https://physics.carleton.ca/clrp/egs_brachy/seed_database_v2}).


\section{Acknowledgements}
This work was supported by the Natural Sciences and Engineering Research
Council of Canada, the Canada Research Chairs program,  the Ministry of
Research and Innovation of Ontario, and a Compute Canada National Resource Allocation. The authors thank Mark Rivard for clarification of source descriptions, as well as Wes Culberson, and Larry DeWerd for providing information on the CivaString sources.


\newpage

\clearpage
\newpage

\section*{References}
\addcontentsline{toc}{section}{\numberline{}References}
\vspace{-1.5cm}
\setlength{\baselineskip}{0.43cm}	




\end{document}